\documentclass[%
preprint,
superscriptaddress,
showpacs,preprintnumbers,
 amsmath,amssymb,
 aps,
prc
]{revtex4-1}
\usepackage{graphicx}
\usepackage{dcolumn}
\usepackage{bm}

\usepackage{subfigure}
\usepackage{mathrsfs}
\usepackage{multirow}

\usepackage[colorlinks,
linkcolor={red!60!black!},
anchorcolor={green!80!black!},
urlcolor={blue!80!black!},
citecolor={blue!50!black!}]{hyperref}
\usepackage[table]{xcolor}

\begin{document}


\title{{\it Ab initio} Gamow in-medium similarity renormalization group with resonance and continuum}

\author{B. S. Hu}
\author{Q. Wu}
\affiliation{School of Physics, and State Key Laboratory of Nuclear Physics and Technology, Peking University, Beijing 100871, China}
\author{Z. H. Sun}
\affiliation{School of Physics, and State Key Laboratory of Nuclear Physics and Technology, Peking University, Beijing 100871, China}
\affiliation{Department of Physics and Astronomy, University of Tennessee, Knoxville, Tennessee 37996, USA}
\affiliation{Physics Division, Oak Ridge National Laboratory, Oak Ridge, Tennessee 37831, USA}
\author{F. R. Xu} \thanks{frxu@pku.edu.cn}
\affiliation{School of Physics, and State Key Laboratory of Nuclear Physics and Technology, Peking University, Beijing 100871, China}


\date{\today}

\begin{abstract}
We have developed a novel {\it ab initio} Gamow in-medium similarity renormalization group (Gamow IMSRG) in the complex-energy Berggren framework. The advanced Gamow IMSRG is capable of describing the resonance and nonresonant continuum properties of weakly-bound and unbound nuclear many-body systems. As test grounds, carbon and oxygen isotopes have been calculated with chiral two- and three-nucleon forces from the effective field theory. Resonant states observed in the neutron-dripline $^{24}$O are well reproduced. The halo structure of the known heaviest Borromean nucleus $^{22}$C is clearly seen by calculating the density distribution, in which continuum $s$-channel plays a crucial role. Further, we predict low-lying resonant excited states in $^{22}$C. The Gamow IMSRG provides tractable {\it ab initio} calculations of weakly-bound and unbound open quantum systems.
\end{abstract}

\maketitle

{\it Introduction.$-$}
Thanks to advanced radioactive beam facilities, loosely-bound and unbound nuclei with extreme neutron-to-proton ratios have been explored in an unprecedented way.
The nuclei belong to the category of open quantum systems in which the coupling to the particle continuum profoundly affects the behavior of the system \cite{OKOLOWICZ2003271,0954-3899-36-1-013101}.
Many novel phenomena have been observed or predicted in the exotic nuclei, such as halos \cite{PhysRevLett.55.2676,RevModPhys.76.215}, genuine intrinsic resonances \cite{PhysRevLett.89.042501,PhysRevLett.89.042502} and new collective modes \cite{PhysRevLett.114.192502,PhysRevC.94.054302}.
To include the continuum effect, several models have been developed, e.g., the continuum shell model (CSM) \cite{BENNACEUR1999289,OKOLOWICZ2003271,PhysRevLett.94.052501}, the Gamow shell model \cite{PhysRevLett.89.042501,PhysRevLett.89.042502}, the complex coupled cluster (CC) \cite{HAGEN2007169,PhysRevLett.108.242501} and the continuum-coupled shell model \cite{10.1093/ptep/ptv125}. 

Current nuclear theory is pursuing {\it ab initio} calculations which are based on realistic nuclear forces and rigorous many-body methods. However, it is always a big challenge to develop an {\it ab initio} method to efficiently describe the continuum. As one of powerful {\it ab initio} renormalizations of interacting Hamiltonians, similarity renormalization group (SRG) was proposed independently by G\l{}azek and Wilson \cite{PhysRevD.48.5863} and by Wegner \cite{AnnPhys.506.77}. Later, Bogner {\it et al.} \cite{PhysRevC.75.061001,PhysRevC.75.051001} applied the SRG method to softening nuclear forces for {\it ab initio} calculations. Recently, the SRG method was developed as a new many-body method in nuclear configuration space,
named in-medium SRG (IMSRG) \cite{PhysRevLett.106.222502,HERGERT2016165}.
The IMSRG can directly give the ground-state properties of closed-shell nuclei \cite{PhysRevLett.106.222502}, and as well be used to derive non-perturbative effective Hamiltonians for the descriptions of excited states or open-shell nuclei \cite{PhysRevC.85.061304,PhysRevLett.113.142501}.
The IMSRG has been developed further, including multi-reference IMSRG \cite{PhysRevLett.110.242501},
IMSRG using an ensemble reference \cite{PhysRevLett.118.032502},
equation-of-motion IMSRG (EOM-IMSRG) \cite{PhysRevC.95.044304,PhysRevC.96.034324}
and IMSRG merging no-core shell model \cite{PhysRevLett.118.152503}.
The IMSRG has become a powerful and predictive {\it ab initio} method.
However, all the existing IMSRG calculations are performed in the harmonic oscillator (HO) or real-energy Hartree-Fock (HF) basis (here the real-energy HF means that the HF approach is performed in the HO basis).
The HO basis is bound and localized, and hence isolated from the environment of unbound scattering states.
The real-energy HF basis also cannot include the continuum. 
It is lacking to include the continuum effect in IMSRG.

The complex-energy Berggren basis provides an efficient framework to treat bound, resonant and nonresonant continuum states, on equal footing \cite{BERGGREN1968265,LIOTTA19961}. Within the Berggren basis, the Gamow shell model \cite{PhysRevLett.89.042501,PhysRevLett.89.042502,PhysRevC.73.064307,PhysRevC.88.044318,SUN2017227} and complex coupled cluster \cite{HAGEN2007169,PhysRevLett.108.242501} have been well developed.
In this paper, we develop the IMSRG in the Berggren basis, and call it the Gamow IMSRG.
As test grounds, we have applied it to oxygen and carbon isotopes.
The recent experiments \cite{PhysRevLett.104.062701,PhysRevC.86.054604,PhysRevLett.109.202503,TOGANO2016412} highlight that $^{22}$C is the heaviest Borromean halo nucleus observed.
The experimental root-mean-squared matter radius of $^{22}$C was deduced to be $3.44\pm0.08$ fm \cite{TOGANO2016412}.
The coupling to continuum should play a role in producing the extended density distribution.
No information has been known experimentally about $^{22}$C excited states which can provide further understanding of halo structure.
Using the Gamow IMSRG, we give a continuum-coupled calculation of the halo $^{22}$C from first principles for the first time.

{\it Gamow Hartree-Fock.$-$}
The intrinsic Hamiltonian of a $A$-nucleon system reads
\begin{eqnarray}
\begin{split}
H=&
\displaystyle\sum_{i=1}^{A} \left(1-\dfrac{1}{A}\right) \dfrac{{\bm{p}_{i}}^{2}}{2m} +
\displaystyle\sum_{i<j}^{A}   \left(V_{\text{NN},ij}-\dfrac{\bm{p}_{i}\cdot\bm{p}_{j} }{mA} \right)
\\&
+\displaystyle\sum_{i<j<k}^{A}{V}_{\text{NNN},ijk},
\label{Hamiltonian}
\end{split}
\end{eqnarray}
where $V_{\text{NN}}$ and $V_{\text{NNN}}$ are the two-nucleon (NN) and three-nucleon (NNN) interactions, respectively. The NN force includes the Coulomb interaction between protons. 
In the present work, we take the optimized chiral NN interaction NNLO$_{\text{opt}}$ \cite{PhysRevLett.110.192502} and also the NNLO$_{\text{sat}}$ which includes the chiral three-nucleon force \cite{PhysRevC.91.051301,PhysRevC.91.044001}. 
The NNLO$_{\text{opt}}$ potential gives the good descriptions of nuclear structures
including binding energies, excitation spectra, dripline positions and the neutron matter equation of state,
without resorting to three-body forces \cite{PhysRevLett.110.192502}. 
The NNLO$_{\text{sat}}$ potential optimizes simultaneously the NN and NNN forces with low-energy nucleon-nucleon scattering data and slected nuclear structure data, specially improving the calculations of nuclear radii \cite{PhysRevC.91.051301}. 

In the Gamow calculations, it is a key step how to choose a proper one-body potential to generate the resonance and continuum Berggren basis.
In many cases, the phenomenological Woods-Saxon potential is used \cite{PhysRevLett.89.042501,PhysRevLett.89.042502,PhysRevLett.120.212502,SUN2017227}.
In the present paper, we use the Gamow Hartree-Fock (GHF) \cite{PhysRevC.73.064307,PhysRevC.88.044318}, with the same chiral potentials, to produce the Berggren single-particle basis. This gives a more self-consistent and {\it ab initio} calculation. 
In detail, we first perform a HF calculation of Hamiltonian (\ref{Hamiltonian}) in the HO basis, giving the HF single-particle states, 
\begin{eqnarray}
|\alpha \rangle=\displaystyle\sum_{p}D_{p\alpha}|p\rangle,
\end{eqnarray}
where $|p\rangle$ indicates the HO basis and the coefficients $D_{p\alpha}$ are determined in the HF diagonalizing. 
After that, we obtain the one-body HF potential $U$ in the HO basis,
\begin{eqnarray}
\begin{split}
\langle p|U|q\rangle
=&\displaystyle\sum_{i=1}^{A}\displaystyle\sum_{rs}\langle pr|V_{\text{NN}}|qs\rangle
D^{\ast}_{ri}D_{si}
\\&+\dfrac{1}{2}\displaystyle\sum_{i,j=1}^{A}\displaystyle\sum_{rstu}\langle prt|V_{\text{NNN}}|qsu\rangle
D^{\ast}_{ri}D_{si}D^{\ast}_{tj}D_{uj}.
\label{uhf}
\end{split}
\end{eqnarray}
The one-body GHF Hamiltonian in the complex-momentum (complex-$k$) space is given by
\begin{eqnarray}
\langle k|h|k'\rangle=(1-\dfrac{1}{A})\dfrac{\hbar^{2}k^{2} }{2m} \delta_{kk'}
+\displaystyle\sum_{pq}\langle p|U|q\rangle \langle k|p \rangle \langle q| k' \rangle,
\label{ghf}
\end{eqnarray}
where $\langle k| p \rangle$ is the HO basis wavefunction $|p\rangle$ expressed in the complex-$k$ space $\langle k|$.
In practical calculations, the momentum is discretized in the contour of the Berggren complex-$k$ plane \cite{SUN2017227}.
Bound, resonant and continuum GHF basis can be obtained by diagonalizing the complex-energy HF Hamiltonian (\ref{ghf}).

{\it Gamow IMSRG.$-$}
Within the GHF basis, the Gamow IMSRG has been performed. The philosophy of SRG is to suppress off-diagonal matrix elements
and drive the original Hamiltonian $H(0)=H$ [given by Eq.(\ref{Hamiltonian})] towards a band- or block-diagonal form
by means of the continuous similarity transformation $U(s)$ with $U(s) \cdot U^{-1}(s)=1$
\cite{PhysRevD.48.5863,AnnPhys.506.77,PhysRevLett.106.222502,HERGERT2016165},
\begin{eqnarray}
H(s)=U(s)H(0)U^{-1}(s).
\end{eqnarray}
The Hamiltonian in the real-momentum space is Hermitian, $H=H^{\dag}$,
therefore the similarity transformation  $U(s)$ can be a unitary transformation in practice,
$U(s) \cdot U^{\dag}(s)=U(s) \cdot U^{-1}(s)=1$, giving
\begin{eqnarray}
H(s)=U(s)H(0)U^{\dag}(s).
\end{eqnarray}
The differential gives the operator flow equation \cite{PhysRevD.48.5863,AnnPhys.506.77},
\begin{eqnarray}
\frac{dH(s)}{ds}=[\eta(s),H(s)],
\label{eq1:imsrg}
\end{eqnarray}
with the anti-Hermitian generator $\eta(s)$,
\begin{eqnarray}
\eta(s)=\dfrac{dU(s)}{ds}U^{\dag}(s)=-\eta^{\dag}(s).
\end{eqnarray}
In the present work, we extend the SRG to the complex-momentum Berggren basis
in which the Hamiltonian becomes complex symmetric, $H=H^{T}$ (here $T$ indicates the transpose) \cite{0954-3899-36-1-013101}.
We perform continuous orthogonal transformation, $U(s) \cdot U^{T}(s)=U(s) \cdot U^{-1}(s)=1$,
to make the $H(0)$ band- \cite{PhysRevD.48.5863} or block-diagonal \cite{PhysRevLett.106.222502}, 
\begin{eqnarray}
H(s)=U(s)H(0)U^{T}(s).
\end{eqnarray}
Correspondingly, the generator appearing in the operator flow equation (\ref{eq1:imsrg}) becomes
\begin{eqnarray}
\label{eta}
\eta(s)=\dfrac{dU(s)}{ds}U^{T}(s)=-\eta^{T}(s).
\end{eqnarray}

If the SRG is implemented in the configuration space of a $A$-body system, it is called the in-medium SRG (IMSRG) \cite{PhysRevLett.106.222502}, as mentioned in the Introduction. 
In the present paper, the IMSRG is developed in the complex-momentum Berggren basis (i.e., the GHF basis described above). We name it the Gamow IMSRG.
The IMSRG itself contains many-body correlations, and can directly give the ground state of a closed-shell nucleus by decoupling the Hamiltonian with the closed shell.
To calculate open-shell nuclei or excited states,
we explore the Gamow IMSRG with the equation of motion (EOM), named Gamow EOM-IMSRG.
The general framework of real-energy EOM-IMSRG can be found in Refs. \cite{PhysRevC.95.044304,PhysRevC.96.034324}. 
In the Gamow EOM-IMSRG calculations, we truncate the excitation operator at a two-particle two-hole level.
The Magnus formulation \cite{PhysRevC.92.034331}
and the White generator $\eta(s)$ \cite{HERGERT2016165} are adopted to decouple the Hamiltonian.
By using the White generator, Eq.~(\ref{eta}) can be satisfied easily. 
Besides, with the White generator, the off-diagonal elements of the Hamiltonian are suppressed with a decay scale $(s-s_{0})$ \cite{HERGERT2016165}.

{\it Calculations and results.$-$}
As mentioned above, the chiral NNLO$_{\text{opt}}$(NN) and  NNLO$_{\text{sat}}$(NN+NNN) have been used in the present Gamow IMSRG calculations. 
The NNLO$_{\text{sat}}$ interaction matrix elements were provided by Oak Ridge group, with 13 major HO shells at $\hbar\Omega= 22$ MeV \cite{PhysRevC.91.051301}.
The NNLO$_{\text{sat}}$ NNN force is fully taken in the HF calculation, while in the IMSRG calculation we take the normal-ordered form of the NNN force which appears finally at a two-body level, containing normal-ordered zero-, one-, and two-body terms of the NNN force \cite{PhysRevC.76.034302,PhysRevLett.109.052501}. In the NNLO$_{\text{opt}}$ calculations, it is a common use of a total 12 major HO shells with $\hbar\Omega= 20$ MeV \cite{PhysRevC.95.014306,PhysRevC.94.014303,1674-1137-41-10-104101,HENDERSON2018468}. 

As a test ground, we have investigated neutron-rich closed-shell oxygen and carbon isotopes. 
Due to the huge computational cost (particularly when the continuum is included), only the neutron $d_{3/2}$ and $s_{1/2}$ partial waves are treated in the resonance and continuum GHF (Berggren) basis, while other neutron channels and all proton channels are handled in the real-energy discrete HF basis (that is obtained in the HO basis) .
Such handling has been adopted in other Gamow many-body calculations, e.g., no core Gamow shell model \cite{PhysRevC.88.044318} and complex coupled cluster \cite{PhysRevLett.108.242501}.
For the $sd$ shell, the neutron $0d_{3/2}$ is a narrow resonant orbit and the $1s_{1/2}$ orbit may have significant effect on the extended spatial distributions of loosely-bound and unbound nuclei.
Therefore, the $d_{3/2}$ and $s_{1/2}$ continua should be explicitly included in the valence model space of the Gamow many-body calculations.

In Ref.~\cite{Morris}, an IMSRG approximation named Magnus(2*) was suggested,
in which a class of undercounted terms in the normal-ordered two-body truncation are restored
by introducing an auxiliary one-body operator with hole-hole and particle-particle excitations.
Magnus(2*) includes more correlations which are from intermediate three-body forces, and brings the two-body truncated IMSRG into an agreement with CCSD method \cite{Morris,jcp141,PhysRevC.95.044304}.
We use the IMSRG Magnus(2*) approximation to evolve the initial many-body Gamow Hamiltonian to be decoupled with the ground state of the closed-shell nucleus.
This gives the ground state and the decoupled Gamow IMSRG Hamiltonian for excited-state calculations using EOM (i.e., Gamow EOM-IMSRG).
With the Gamow IMSRG Hamiltonian obtained thus, we perform the EOM calculation for excited states.
Only hole-hole excitations are considered in the auxiliary one-body operator of the Magnus(2*) approximation.
In Ref.~\cite{EVANGELISTA201227}, it has been shown that the contribution of particle-particle excitations is significantly smaller than the hole-hole contribution.

Although Hamiltonian (\ref{Hamiltonian}) is intrinsic, the IMSRG wave function is expressed in the laboratory coordinate.
One might think of the correction from the center-of-mass (CoM) motion. 
In the HO basis, the CoM motion can be treated using the Lawson method \cite{lawson}.
In the real-energy HF basis, an approximation similar to the Lawson method was suggested in the CC and IMSRG calculations \cite{PhysRevLett.103.062503,HERGERT2016165}.
Unfortunately, the method cannot be used in the complex-energy Berggren basis,
due to the fact that the $R^2$ matrix elements ($R$ is the CoM position) cannot be regularized in resonance and continuum states which are not square integrable.

\begin{figure}
\centering
\setlength{\abovecaptionskip}{6pt}
\setlength{\belowcaptionskip}{6pt}
\includegraphics[scale=0.6]{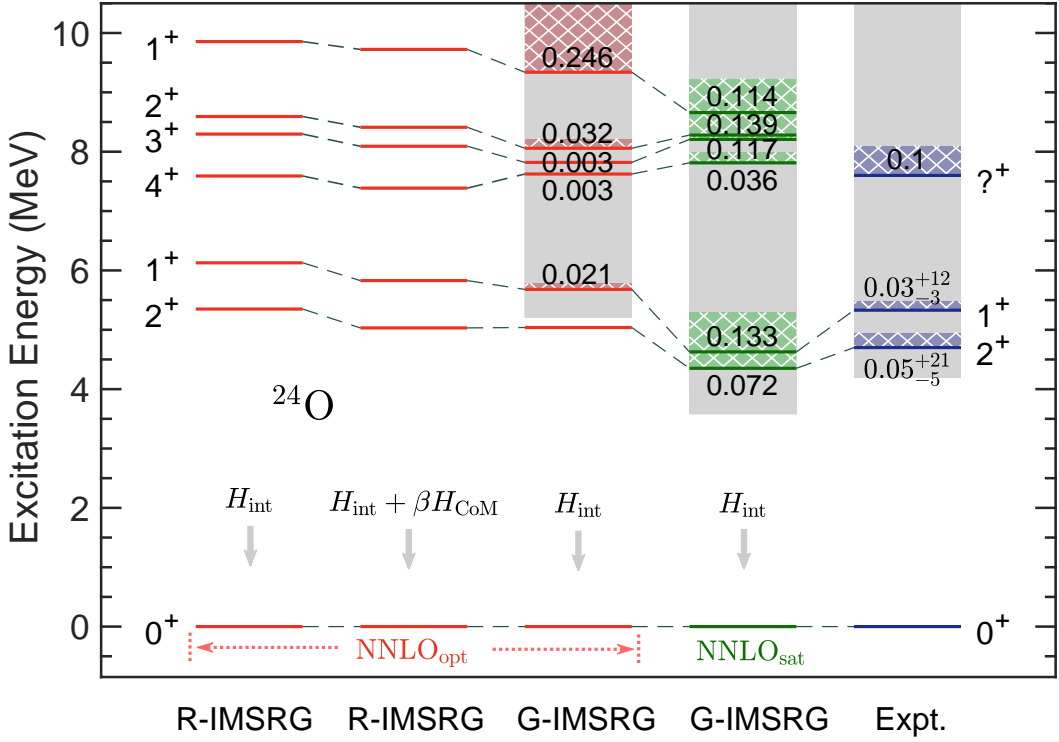}
\caption{$^{24}$O spectrum calculated with NNLO$_{\rm{opt}}$(NN) and NNLO$_{\rm{sat}}$(NN+NNN).
The left two columns give the real-energy EOM-IMSRG calculations (indicated by R-IMSRG) without and with the CoM treatment $\beta H_{\text{CoM}}$. We take the multiplier $\beta=5$.
The right three columns show the Gamow EOM-IMSRG calculations (indicated by G-IMSRG) and data \cite{HOFFMAN200917,PhysRevC.83.031303}. 
Resonant states are indicated by shading, and their widths (in MeV) are given by the numbers nearby. The gray shading means the continuum above the particle emission threshold. $H_{\text int}$ indicates the intrinsic Hamiltonian (\ref{Hamiltonian}).}
\label{o24}
\end{figure}

In the previous work \cite{SUN2017227}, we have discussed that the CoM effect with an intrinsic Hamiltonian is small for low-lying states.
In the present paper, we use the approximation suggested in Refs. \cite{PhysRevLett.103.062503,HERGERT2016165} to estimate the CoM effect in the IMSRG calculation.
In Figure~\ref{o24}, we show the real-energy EOM-IMSRG calculations (here ``real-energy" indicates that the calculation is performed in the real-energy HF basis) without and with the multiplied CoM term, $\beta H_{\rm{CoM}}=\beta \left( \dfrac{\bm{P^{2}}}{2mA}+\dfrac{1}{2}mA \tilde{\Omega}^{2}\bm{R}^{2}-\dfrac{3}{2}\hbar\tilde{\Omega} \right)$.
The value of the CoM vibration frequency $\tilde\Omega$ can be different from the frequency $\Omega$ of the HO basis
in which the HF equation is solved \cite{PhysRevLett.103.062503,HERGERT2016165}.
The `best' $\tilde\Omega$ value can be determined by minimizing the $H_{\text{CoM}}(\tilde\Omega)$ expectation value of the state (e.g., the ground state), though the total energy of a state should not be sensitive to the $\tilde\Omega$ value \cite{PhysRevLett.103.062503}. We find that the minimized $H_{\text{CoM}}(\tilde\Omega)$ expectation values are approximately zero, when $\tilde\Omega\approx$ 14.0 and 12.6 MeV in the ground states of $^{24}$O and $^{22}$C, respectively. 
It is seen that, in Figure~{\ref{o24}}, the CoM effect is small. 


The detailed Gamow EOM-IMSRG calculations for $^{24}$O are shown in Figure~\ref{o24}, with the chiral NNLO$_{\rm opt}$(NN) \cite{PhysRevLett.110.192502} and NNLO$_{\text{sat}}$(NN+NNN) \cite{PhysRevC.91.051301,PhysRevC.91.044001} interactions. 
The converged calculations are independent of the choice of the contour for the Berggren partial waves. We have tested that for the $sd$ shell 30 Gauss-Legendre mesh points for each of the continuum channels is sufficient to provide the converged calculations of the energies of nuclear states (including binding energies). 
The present calculations reproduce the experimental excitation energies and resonances of the observed states.
The high excitation energy of the first $2^+$ state in $^{24}$O supports
the shell closure at $N$=16 in the oxygen chain.
The calculations predict three resonant states around the excitation energies of $\sim 8$ MeV with $J^{\pi}=2^{+}, 3^{+}$, $4^{+}$, corresponding to the not-yet-clear experimental states around 7.6 MeV \cite{PhysRevC.83.031303}.
This prediction is consistent with the complex coupled-cluster (CC) calculation with a schematic three-nucleon force \cite{PhysRevLett.108.242501}.

\begin{figure}
\centering
\setlength{\abovecaptionskip}{6pt}
\setlength{\belowcaptionskip}{6pt}
\includegraphics[scale=0.6]{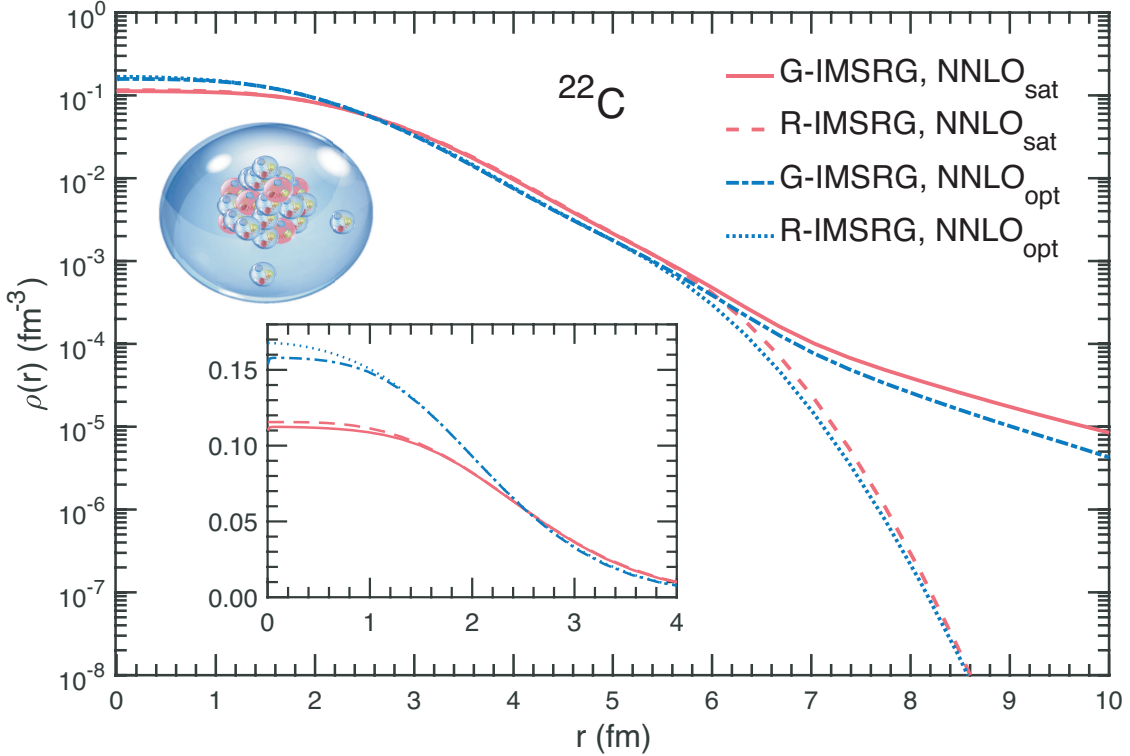}
\caption{Calculated $^{22}$C ground-state densities displayed in the logarithm scale.
R-IMSRG indicates the real-energy IMSRG calculation,
while G-IMSRG is the Gamow IMSRG calculation.
The inset details the densities in the central region of the nucleus with the standard scale.}
\label{density}
\end{figure}
The Borromean halo $^{22}$C is a challenging nucleus for many theoretical calculations \cite{ACHARYA2013196,SUZUKI2016199,SUN2018530}.
Our Gamow Hartree-Fock calculation gives that the neutron $\nu 1s_{1/2}$ orbital is weakly bound.  
The two-neutron configuration $[\nu 1s_{1/2}]^{2}$ is responsible for the halo formation \cite{PhysRevLett.104.062701,PhysRevC.86.054604,PhysRevLett.109.202503,TOGANO2016412}.
We have performed IMSRG calculations for $^{22}$C. 
Figure~\ref{density} shows the ground-state density obtained with an effective density operator. 
The operator is derived self-consistently using the Baker-Cambell-Hausdorff formulation \cite{PhysRevC.92.034331} within the Magnus framework of IMSRG.
We see that the Gamow IMSRG calculation gives a long tail in the density distribution, supporting the halo structure. Note that more than 30 Gauss-Legendre mesh points would be needed to obtain a converged and smooth tail in the density distribution of the halo nucleus. In $^{22}$C, we find that 36 mesh points can lead to a converged and smooth tail of the density distribution. Calculations with more mesh points cost much more computer time. 

To see the effect from the continuum in $^{22}$C, we have analyzed the role of the $s$-channel. Two types of the Gamow IMSRG calculations have been performed: 
(i) with discrete $s$-states which are obtained in the real-energy HF calculation,
and (ii) with the Berggren $s$-states that are obtained in the complex-energy GHF calculation.
In both calculations, the neutron $d_{3/2}$ channel remains in the GHF basis.
The Gamow IMSRG calculation with adopting the discrete real-energy HF $s$-states gives a matter radius of 2.798 fm for the $^{22}$C ground state, while the calculation with taking the continuum $s$-wave gives a larger radius of 2.928 fm.
The similar calculations with NNLO$_{\text{sat}}$(NN+NNN) have also been performed.  
The obtained radius is 2.983 fm with the real-energy discrete $s$-states and 3.139 fm with continuum GHF $s$-wave.
The experimental estimated  matter radius was $5.4\pm0.9$ fm in an earlier work \cite{PhysRevLett.104.062701} and it is $3.44\pm0.08$ fm \cite{TOGANO2016412} and 3.38$\pm$0.10 fm \cite{PhysRevC.97.054614} in the later works.
We see that the continuum $s$-wave plays an important role in the calculation of the radius and the halo structure.
The NNLO$_{\text{opt}}$(NN) itself underestimates the radii of carbon isotopes \cite{PhysRevLett.117.102501},
while the NNLO$_{\rm sat}$(NN+NNN) give good descriptions of radii \cite{PhysRevC.91.051301,PhysRevLett.117.102501}. 
A recent calculation of the relativistic mean-field model with continuum gives a radius of 3.25 fm \cite{SUN2018530}. 

\begin{figure}
\centering
\setlength{\abovecaptionskip}{6pt}
\setlength{\belowcaptionskip}{6pt}
\includegraphics[scale=0.6]{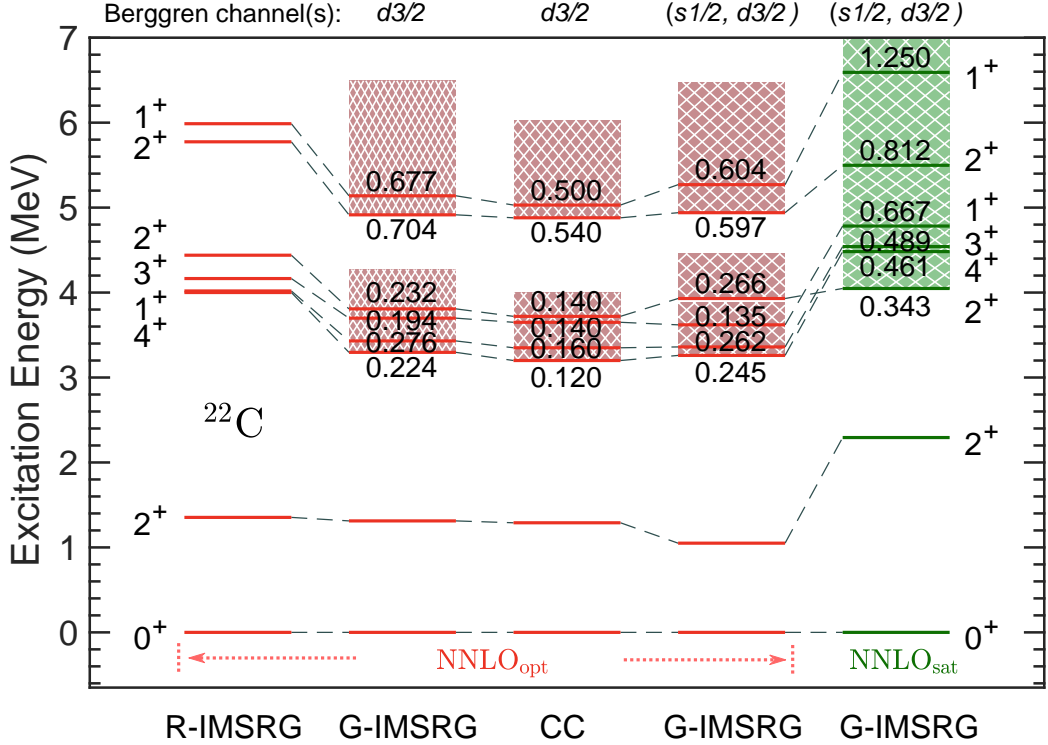}
\caption{Excited states in $^{22}$C predicted by Gamow EOM-IMSRG with NNLO$_{\rm{opt}}$(NN) and NNLO$_{\rm{sat}}$(NN+NNN), compared with the complex CC calculation \cite{Hagen}. The channel(s) given on the top of the panel indicates that the partial waves are treated in the resonance and continuum Berggren representation. Other labels are similar to in Figure~\ref{o24}.}
\label{c22}
\end{figure}
There have been no experimental data available for excited states in $^{22}$C.
Figure~\ref{c22} gives the Gamow EOM-IMSRG predictions for possible low-lying states.
The results are benchmarked with the complex CC calculations \cite{Hagen}. We see that the two types of calculations are consistent with each other.
The first $2^{+}$ state is bound in the present and CC  calculations.
We find that the $2_1^+$ state is dominated by the proton $1p$-$1h$ excitation from the $\pi 0p_{3/2}$ hole to $\pi 0p_{1/2}$ particle orbits.
The proton $2^+$ excited state is lower in energy than the neutron $2^+$ state that has been calculated by the real-energy shell model with the $^{14}$C core \cite{PhysRevC.81.064303,PhysRevLett.113.142502}. 
The present real-energy IMSRG (see Figure~\ref{c22}) gives a similar neutron $2^+$ energy to in Refs. \cite{PhysRevC.81.064303,PhysRevLett.113.142502}, while the Gamow IMSRG and complex CC predict a lower neutron $2^+$ energy by about 0.6 MeV. 
In fact, there are superposed resonant states with $J^{\pi}=1^{+}$, $2^+$, $3^+$, $4^{+}$ at energies $\sim3.5-4.0$ MeV and widths $\sim0.15-0.25$ MeV. 
The NNLO$_{\text{sat}}$(NN+NNN) calculations give slightly higher excitation energies and wider resonant widths for the superposed states, as shown in Figure~\ref{c22}.   
The resonances are dominated by neutron $1p$-$1h$ excitations from $\nu 0d_{5/2}$ hole to $\nu 0d_{3/2}$ particle.

{\it Summary.$-$}
We have developed the {\it ab initio} Gamow in-medium similarity renormalization group (Gamow IMSRG) which includes the continuum via the complex-energy Berggren basis obtained by the Gamow Hartree-Fock with chiral interactions. Using the Gamow IMSRG, the continuum-coupled Hamiltonian of a closed-shell nucleus is decoupled first with the ground-state configuration, which gives meanwhile the ground-state property of the nucleus. With the decoupled IMSRG Hamiltonian, we perform the equation of motion (EOM) calculation to obtain excited states, which we call Gamow EOM-IMSRG. The method provides a unified description of bound, resonant and continuum states of nuclei. As a test ground, we have calculated the neutron-dripline nucleus $^{24}$O in which resonant states have been observed experimentally. The present calculations reproduce well the experimental observations. $^{22}$C has also been investigated, giving the known halo structure. Low-lying resonant states in  $^{22}$C are predicted, providing useful information for future experiments. The calculation shows that the continuum $s$-wave leads to a large spatial extension of the Borromean halo nucleus $^{22}$C.

\begin{acknowledgments}
We are grateful to G.R. Jansen and T. Papenbrock for producing the NNLO$_{\rm sat}$ matrix elements for us, and to G. Hagen for providing the coupled-cluster calculations for our benchmarking.
Valuable discussions with Simin Wang, N. Michel, M. P\l{}oszajczak, G. Hagen
and C.W. Johnson are gratefully acknowledged.
This work has been supported by
the National Key R${\&}$D Program of China under Grant No. 2018YFA0404401;
the National Natural Science Foundation of China under Grants No. 11835001, No. 11575007 and No. 11847203;
China Postdoctoral Science Foundation under Grant No. 2018M630018;
and the CUSTIPEN (China-U.S. Theory Institute for Physics with Exotic Nuclei) funded by the U.S.  Department of Energy,
Office of Science under Grant No. DE-SC0009971.
We acknowledge the High-performance Computing Platform of Peking University for providing computational resources.
\end{acknowledgments}

\bibliography{references}

\end{document}